\documentstyle[twocolumn,prb,aps,epsfig,amssym]{revtex}

\begin{document}
\draft
\preprint{}

\newcommand{\1}{{\bf \scriptstyle 1}\!\!{1}}
\newcommand{\I}{{\rm i}}
\newcommand{\p}{\partial}
\newcommand{\D}{^{\dagger}}
\newcommand{\bx}{{\bf x}}
\newcommand{\bk}{{\bf k}}
\newcommand{\bv}{{\bf v}}
\newcommand{\bp}{{\bf p}}
\newcommand{\bu}{{\bf u}}
\newcommand{\bA}{{\bf A}}
\newcommand{\bB}{{\bf B}}
\newcommand{\bK}{{\bf K}}
\newcommand{\bL}{{\bf L}}
\newcommand{\bP}{{\bf P}}
\newcommand{\bQ}{{\bf Q}}
\newcommand{\bS}{{\bf S}}
\newcommand{\bH}{{\bf H}}
\newcommand{\balpha}{\mbox{\boldmath $\alpha$}}
\newcommand{\bsigma}{\mbox{\boldmath $\sigma$}}
\newcommand{\bSigma}{\mbox{\boldmath $\Sigma$}}
\newcommand{\bomega}{\mbox{\boldmath $\omega$}}
\newcommand{\bpi}{\mbox{\boldmath $\pi$}}
\newcommand{\bphi}{\mbox{\boldmath $\phi$}}
\newcommand{\bnabla}{\mbox{\boldmath $\nabla$}}
\newcommand{\bmu}{\mbox{\boldmath $\mu$}}
\newcommand{\bepsilon}{\mbox{\boldmath $\epsilon$}}

\newcommand{\iLambda}{{\it \Lambda}}
\newcommand{\cL}{{\cal L}}
\newcommand{\cH}{{\cal H}}
\newcommand{\cU}{{\cal U}}
\newcommand{\cT}{{\cal T}}

\newcommand{\be}{\begin{equation}}
\newcommand{\ee}{\end{equation}}
\newcommand{\bea}{\begin{eqnarray}}
\newcommand{\eea}{\end{eqnarray}}
\newcommand{\beqa}{\begin{eqnarray*}}
\newcommand{\eeqa}{\end{eqnarray*}}
\newcommand{\nn}{\nonumber}
\newcommand{\DD}{\displaystyle}

\newcommand{\ba}{\left[\begin{array}{c}}
\newcommand{\baa}{\left[\begin{array}{cc}}
\newcommand{\baaa}{\left[\begin{array}{ccc}}
\newcommand{\baaaa}{\left[\begin{array}{cccc}}
\newcommand{\ea}{\end{array}\right]}

\twocolumn[
\hsize\textwidth\columnwidth\hsize\csname
@twocolumnfalse\endcsname

\title{Incoherent Zener tunneling and its application to molecular magnets}

\author{Michael N.~Leuenberger\cite{email1} and Daniel Loss\cite{email2}}
\address{Department of Physics and Astronomy, University of Basel \\
Klingelbergstrasse 82, 4056 Basel, Switzerland}

\maketitle

\begin{abstract}
We generalize the Landau-Zener theory of coherent tunneling transitions by taking thermal relaxation into account. The evaluation of a generalized master equation containing a dynamic tunneling rate that includes the interaction between the relevant system and its environment leads to an incoherent Zener transition probability with an exponent that is twice as large as the one of the coherent Zener probability in the limit $T\rightarrow 0$. We apply our results to molecular clusters, in particular to recent measurements of the tunneling transition of spins in Fe$_8$ crystals performed by Wernsdorfer and Sessoli \protect{[Science {\bf 284}, 133 (1999)]}.
\end{abstract}

]
\narrowtext

The adiabatic transition in a two-level system $\left\{\left|m\right>,\left|m'\right>\right\}$ with energy level crossing is described by the Landau-Zener transition probability\cite{Zener}
\be
P_{\rm coh}=1-\exp\left(-\frac{\pi E_{mm'}^2}{2\hbar\left|\frac{d}{dt}(\varepsilon_m-\varepsilon_{m'})\right|}\right),
\label{zener}
\ee
where $\varepsilon_m$ ($\varepsilon_{m'}\approx\varepsilon_m$) is the energy of the state $\left|m\right>$ ($\left|m'\right>$), and $E_{mm'}$ is given by the coupling between these states. Equation (\ref{zener}) is obtained directly from the Schr\"odinger equation with the Hamiltonian\cite{Zener}
\be
\cH=\baa \varepsilon_m & E_{mm'}/2 \\ E_{mm'}/2 & \varepsilon_{m'} \ea.
\ee
If the total Hamiltonian $\cH_{\rm tot}$ comprising $\cH$ forms a potential barrier --- as we shall assume from now on --- $E_{mm'}$ corresponds to the tunnel splitting energy, $P_{\rm coh}$ to the coherent Zener tunneling probability, and the eigenstates of $\cH$ are delocalized as long as $\varepsilon_m-\varepsilon_{m'}\lesssim E_{mm'}$.
As there is a large amount of potential barriers in physical systems, Eq.~(\ref{zener}) has become an important tool for studying tunneling transitions.\cite{Tool1,Tool2,Tool3,Tool4} It must be noted that all quantum systems to which the Zener model\cite{Zener} is applicable can be described by {\it pure} states and their {\it coherent} time evolution. It is the aim of the present work to generalize the Zener theory in the sense that we take also the {\it incoherent} evolution of {\it mixed} states into account (see also Refs. \onlinecite{Tool1} and \onlinecite{DZener1,DZener2,DZener3,DZener4,DZener5} for a comparison). In order to provide a clear description of our generally valid theory, we give the derivation of the incoherent Zener tunneling probability $P_{\rm inc}$ (see Fig.~\ref{crossing}) in the framework of spin tunneling in molecular magnets, which has become a highly attractive research field in the past few years since several experiments
revealed mesoscopically observable quantum phenomena in molecular clusters, such as Mn$_{12}$-acetate (Mn$_{12}$) (Refs.~\onlinecite{Paulsen,Paulsen et al,Sessoli,Novak et al,Friedman,Thomas}) and Fe$_8$-triazacyclononane (Fe$_8$).\cite{Barra,Sangregorio,Ohm,Wernsdorfer,Wernsdorfer2} In particular, we will show that our theory presented in this work is in good agreement with recent measurements of $P_{\rm inc}$ as a function of the external transversal field $H_x$ for various temperatures in Fe$_8$.\cite{Wernsdorfer,Wernsdorfer2}

\begin{figure}[b]
  \begin{center}
    \leavevmode
\epsfxsize=8.5cm
\epsffile{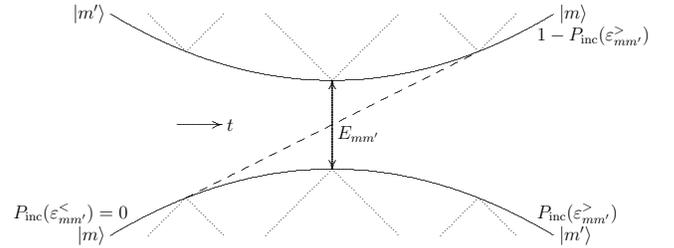}
  \end{center}
\caption{Energy level crossing diagram for incoherent Zener transitions. Dotted lines: transitions due to interaction with environment, leading to a linewidth $\gamma_{mm'}$. Variables are explained in the text.}
\label{crossing}
\end{figure}

We proceed now from the assumption that the range over which $\varepsilon_{mm'}(t)=\varepsilon_m-\varepsilon_{m'}$ is swept, defined by the boundaries $\varepsilon_{mm'}^<:=\min_t\{\varepsilon_{mm'}\}$ and $\varepsilon_{mm'}^>:=\max_t\{\varepsilon_{mm'}\}$, is much larger than $E_{mm'}$ and the decoherence rate $\hbar\gamma_{mm'}$ (see below and Fig.~\ref{crossing}). In addition, we restrict the evolution of our system to times $t$ that are much longer than the decoherence time $\tau_{\rm d}=1/\gamma_{mm'}$. In this case, tunneling transitions between pairs of degenerate excited states are incoherent. This tunneling is only observable if the temperature $T$ is kept well below the activation energy of the potential barrier. Accordingly, we are interested only in times $t$ that are larger than the relaxation times of the excited states. Thus, we can apply our formalism presented in Ref.~\onlinecite{LeuenbergerLoss}, which treats incoherent tunneling between pairwise degenerate states within a single spin system, the decoherence of which is due to the interaction with its environment.
We showed in Ref.~\onlinecite{LeuenbergerLoss} that one can reduce the generalized master equation comprising off-diagonal elements of the density matrix $\rho$ to a complete master equation that consists only of the diagonal elements, i.e.,
\be
\dot{\rho}_m =-W_m\rho_m+\!\!\!\sum_{n\ne m,m'}\!\!\!
W_{mn}\rho_n\,+\Gamma_m^{m'}\left(\rho_{m'}-\rho_m\right),
\label{master}
\ee
where
\be
\Gamma_m^{m'}(t)=\frac{E_{mm'}^2}{2}\frac{\gamma_{mm'}}{\varepsilon_{mm'}^2(t)
+\hbar^2\gamma_{mm'}^2}
\label{lorentzian}
\ee
is the incoherent tunneling rate from $\left|m\right>$ to $\left|m'\right>$, which, in contrast to Ref.~\onlinecite{LeuenbergerLoss}, is assumed now to be time dependent (see below for range of validity). In Eqs.~(\ref{master}) and (\ref{lorentzian}) we have made use of the abbreviations $\gamma_{mm'}=(W_m+W_{m'})/2$ and $W_m=\sum_nW_{nm}$, where $W_{nm}$ denotes the approximately time-independent transition rate from $\left|m\right>$ to $\left|n\right>$, which can be obtained via Fermi's golden rule.\cite{LeuenbergerLoss} 
For straightforward calculation of $E_{mm'}$ it is useful to note that our generalized tunnel splitting formula\cite{LeuenbergerLoss}
\be
E_{mm'}=2\left|\sum\limits_{m_1,\ldots,m_N \atop m_i\ne m,m'}
\frac{V_{m,m_1}}{\varepsilon_m-\varepsilon_{m_1}}\prod\limits_{i=1}^{N-1}
\frac{V_{m_i,m_{i+1}}}{\varepsilon_m-\varepsilon_{m_{i+1}}}V_{m_N,m'}\right|
\label{gsplitting}
\ee
can be represented by the graphs shown in Fig.~\ref{feynman}. $V_{m_i,m_j}$ denote off-diagonal matrix elements of the total Hamiltonian $\cH_{\rm tot}$. 

\begin{figure}[htb]
  \begin{center}
    \leavevmode
\epsfxsize=4cm
\epsffile{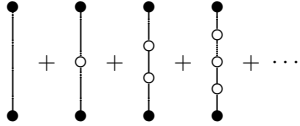}
  \end{center}
\caption{Tunnel splitting energy $E_{mm'}$. The open circles $\circ$ correspond to the states $\left|m_i\right>$, $i=1,\ldots,N$, the solid circles $\bullet$ to $\left|m\right>$, $\left|m'\right>$, and the lines to the matrix elements $V_{m_i,m_j}$.}
\label{feynman}
\end{figure}

First we solve Eq.~(\ref{master}) in the unbiased case --- corresponding to $n=0$ (see below) --- where the ground states $\left|s\right>$, $\left|-s\right>$ and the excited states $\left|m\right>$, $\left|-m\right>$, $m\in\left[[s]-s+1,s-1\right]$ of our spin system with spin $s$ are pairwise degenerate. In addition, we assume that the excited states are already in their stationary state, i.e., $\dot{\rho}_m=0$ $\forall m\ne s,-s$. This implies that $\varepsilon_{mm'}(t)$ in Eq.~(\ref{lorentzian}) must be changed within a time that is much smaller (adiabatic approximation) or much greater (sudden approximation) than the relaxation times of the excited states, which are of the order of $1/W_m$. 
Proceeding as in Ref.~\onlinecite{LeuenbergerLoss} (Sec.~V~A) we obtain from Eq.~(\ref{master})
\be
1-P_{\rm inc}\equiv\Delta\rho(t)=\exp\left\{-\int_{t_0}^t dt'\;\Gamma_{\rm tot}(t')\right\},
\label{exact_solution}
\ee
where we have defined the quantity $\Delta\rho(t)=\rho_{s}-\rho_{-s}$, which satisfies the initial condition $\Delta\rho(t=t_0)=1$, and thus $P_{\rm inc}(t=t_0)=0$. In distinction to Eq.~(\ref{zener}), we call $P_{\rm inc}$ the incoherent Zener transition probability. The total time-dependent relaxation rate is given by
$\Gamma_{\rm tot}=2\left[\Gamma_{s}^{-s}+\Gamma_{\rm th}\right]$,
where the thermal rate $\Gamma_{\rm th}$, which determines the incoherent relaxation via the excited states, is evaluated by means of relaxation diagrams,\cite{LeuenbergerLoss} such as shown in Fig.~\ref{diagram}. For example, if we allow only for thermal transitions with $\Delta m=1$, we obtain
$\Gamma_{\rm th}=f_{s-1}\circ f_{s-2}\circ\cdots\circ f_{[s]/s}(0)$.
This continued fraction is recursively defined by
\[
f_m\circ g:=\frac{b_m}{\frac{2}{W_{m+1,m}}+\frac{1}{\Gamma_{m}^{-m}+g}},\quad 
b_m=e^{-\beta(\varepsilon_m-\varepsilon_s)}.
\]

\begin{figure}[htb]
  \begin{center}
    \leavevmode
\epsfxsize=6cm
\epsffile{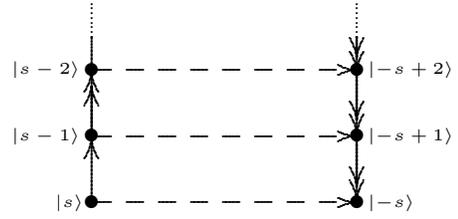}
  \end{center}
\caption{Unbiased ($n=0$) relaxation diagram of a spin system with spin $\bS$ and symmetric anisotropy barrier $\cH_{\rm a}=-AS_z^2$. The solid lines correspond to thermal transitions with $\Delta m\in\left[1,2s\right]$, double arrows indicating that there is more than one incoming rate, and the dashed lines represent tunneling transitions.}
\label{diagram}
\end{figure}

Assuming linear time dependence, i.e., $\varepsilon_{mm'}(t)=\alpha_m^{m'} t$, in the transition region,\cite{Zener} and with $\left|\varepsilon_{mm'}^{<,>}\right|\gg\hbar\gamma_{mm'}$ we obtain from Eq.~(\ref{exact_solution})  
\bea
\Delta\rho
& = & \exp\left\{-\frac{2E_{s,-s}^2}{\hbar\alpha_s^{-s}}
\arctan\left(\frac{\alpha_s^{-s}}{\hbar\gamma_{s,-s}}t\right)
-\int_{-t}^t dt'\;\Gamma_{\rm th}\right\} \nn\\
& \approx & \exp\left\{-\frac{\pi E_{s,-s}^2}{\hbar\alpha_s^{-s}}
-\int_{-t}^t dt'\;\Gamma_{\rm th}\right\},
\label{rate_linear}
\eea
where we have set $t_0=-t$.
In the low-temperature limit $T\rightarrow 0$ the excited states are not populated anymore and thus $\Gamma_{\rm th}$, which consists of intermediate rates that are weighted by Boltzmann factors $b_m$,\cite{LeuenbergerLoss} vanishes.
Consequently, 
Eq.~(\ref{rate_linear}) simplifies to
\be
\Delta\rho=\exp\left\{-\frac{\pi E_{s,-s}^2}{\hbar\alpha_s^{-s}}\right\}=
\exp\left\{-\frac{\pi E_{s,-s}^2}{\hbar\left|\dot{\varepsilon}_{s,-s}(0)\right|}\right\},
\label{rate_linear_limit}
\ee
where the second expression is more general and can also be obtained directly from Eq.~(\ref{lorentzian}) by reducing 
\be
2\hbar\Gamma_s^{-s}\stackrel{\gamma_{s,-s}\rightarrow 0}{\longrightarrow}
E_{s,-s}^2\pi\delta(\varepsilon_{s,-s})=
\frac{E_{s,-s}^2\pi\delta(t)}{\left|\dot{\varepsilon}_{s,-s}(0)\right|}.
\ee
We note that the exponent in Eq.~(\ref{rate_linear_limit}) differs by a factor of 2 from the Zener exponent in Eq.~(\ref{zener}). This is not surprising since $\Gamma_{\rm tot}$ is the relaxation rate of $\Delta\rho$, where both $\rho_{s}$ and $\rho_{-s}$ are changed in time by the same amount, and {\it not} an escape rate like in the cases of coherent Zener transition and $\alpha$ decay, where only the population of the inital state is changed in time.
Note that Eq.~(\ref{rate_linear_limit}) implies $P_{\rm inc}=1$ for $\left|\dot{\varepsilon}_{s,-s}(0)\right|\rightarrow 0$ (adiabatic limit) and $P_{\rm inc}=0$ for $\left|\dot{\varepsilon}_{s,-s}(0)\right|\rightarrow\infty$ (sudden limit).

Instead of the linear time dependence $\varepsilon_{s,-s}=\alpha_s^{-s} t$, one can consider oscillations of the form $\varepsilon_{s,-s}=a\sin\omega t$. Neglecting $\Gamma_{\rm th}$ in Eq.~(\ref{rate_linear}) in the limit $T\rightarrow 0$ we get
\be
\Delta\rho
=\Delta\rho(t_0)\exp\left\{-\frac{E_{s,-s}^2}{\hbar\omega \eta}
\arctan\left[\frac{\eta}{\hbar\gamma_{s,-s}}\tan(\omega t)\right]\right\},
\label{rate_sinus}
\ee
where we have set $\eta=\sqrt{a^2+\hbar^2\gamma_{s,-s}^2}$.
Integrating $\Gamma_{\rm tot}$ from $t_0=-\pi/2\omega$ to $t=\pi/2\omega$ we obtain
$\Delta\rho=\exp\left\{-\pi E_{s,-s}^2/\hbar\omega\eta\right\}$.
In comparison to Eq.~(\ref{rate_linear_limit}) we get here the extra factor $\eta$, which provides an experimentally exploitable dependence on $\gamma_{s,-s}$, provided that $a\lesssim\hbar\gamma_{s,-s}$.

If we apply a bias to our system in such a way that our states are tuned to other resonances (see Fig.~\ref{diagram_biased}), e.g., $n=1,2,\ldots$ (see below), the total relaxation rate is changed to
$\Gamma_{\rm tot}^{\rm bias}\approx 2/\left[ 1/(\Gamma_{s}^{-s+1}+\Gamma_{\rm th}^{\rm bias})+1/W_{-s,-s+1}\right]$,
where the approximation comes from the fact that there can be other nonvanishing thermal rates $W_{-s,n}$ with initial state $\left|n\right>$, $n\ne -s+1$, and final state $\left|-s\right>$. $\Gamma_{\rm th}^{\rm bias}$ determines the relaxation through the states that have higher energy than $\left|s\right>$.

\begin{figure}[htb]
  \begin{center}
    \leavevmode
\epsfxsize=5cm
\epsffile{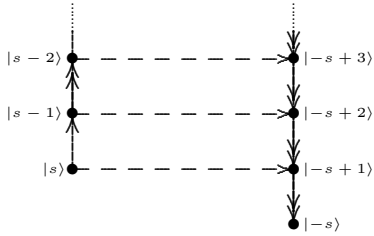}
  \end{center}
\caption{Example of a biased relaxation diagram of a spin system with spin $S_z$ and asymmetric barrier $\cH_{\rm a}+\cH_{\rm Z}$ for the resonance condition $n=1$. The diagram is explained in Fig.~\protect\ref{diagram}.}
\label{diagram_biased}
\end{figure}

In the second part of this paper, we apply our theory to recent experiments,\cite{Wernsdorfer,Wernsdorfer2} which measured quantum oscillations of the tunnel splitting $E_{mm'}(H_x)$ in Fe$_8$ as a function of an externally applied transversal magnetic field $H_x$. In a coherent spin-state path integral approach these quantum oscillations and their associated spin parity effects can be viewed as a result of interfering Berry phases carried by spin tunneling paths of opposite winding,\cite{Loss} which are modified in the presence of a field $H_x$.\cite{Garg} However, for a quantitative analysis of the Fe$_8$ data\cite{Wernsdorfer,Wernsdorfer2} the operator formalism presented here proves to be more useful than the path integral approach.  

In accordance with earlier work\cite{Barra,Sangregorio,Ohm,Wernsdorfer,Wernsdorfer2,Caciuffo} we use a single-spin Hamiltonian $\cH=\cH_{\rm a}+\cH_{\rm T}+\cH_{\rm Z}+\cH_{\rm sp}$ that describes sufficiently well the behavior of the giant spin $\bS$ with $s=10$ of a Fe$_8$ cluster.
It turns out that our theory is in optimal agreement with experiments\cite{Wernsdorfer,Wernsdorfer2} if we choose the easy-axis Hamiltonian to be
$\cH_{\rm a}=-AS_z^2$,
with anisotropy constant $A/k_B=0.275$ K,\cite{Barra} and the in-plane Hamiltonian to be
\bea
\cH_{\rm T} & = & \frac{1}{2}\sum\limits_{n=1}^4 B_{2n}\left(S_+^{2n}+S_-^{2n}\right) \nn\\
& & +\left.\frac{1}{2}g\mu_BH\sin\vartheta\left(e^{-i\varphi}S_++e^{i\varphi}S_-\right)\right.,
\label{H_T}
\eea
with the anisotropy constants $B_2/k_B=0.046$ K,\cite{Barra,Caciuffo} $B_4/k_B=-6.0\times 10^{-5}$ K, $B_6/k_B=2.0\times 10^{-8}$ K, and $B_8/k_B=2.0\times 10^{-11}$ K. Besides the $B_4$ term introduced in Refs.~\onlinecite{Wernsdorfer} and \onlinecite{Wernsdorfer2} to obtain the desired period, we find that the $B_6$ term is necessary to achieve the desired tunneling amplitude in Fig.~\ref{berry}, while the $B_8$ term is responsible for the minimum at $H_x\approx 1.4$ T. The Zeeman coupling $g\mu_B\bS\cdot\bH$ has been divided into a longitudinal part $\cH_{\rm Z}=g\mu_B H_zS_z$ and a transversal part, being the second term in Eq.~(\ref{H_T}), where $H$ is the magnitude of the external magnetic field $\bH$, and $\vartheta$ and $\varphi$ define the spherical angles.
According to Eq.~(\ref{gsplitting}), $\cH_{\rm T}$ induces tunneling between pairwise degenerate $S_z$ eigenstates $\left|m\right>$, $-s\le m\le s$, of $\cH_{\rm a}+\cH_{\rm Z}$, with eigenvalues $\varepsilon_m$. The resonance condition for such degeneracies, i.e., $\varepsilon_m=\varepsilon_{m'}$, leads to the resonance field $H_z^{mm'}=nA/g\mu_B$, $n=m+m'$.
As can be seen in Fig.~\ref{berry}, our analytic formula (\ref{gsplitting}) for the resulting tunnel splitting $E_{mm'}(H_x)$ is in reasonable agreement with the exact diagonalization of $\cH_{\rm a}+\cH_{\rm T}+\cH_{\rm Z}$, which provides a good fit of the data in Ref.\onlinecite{Wernsdorfer}.

\begin{figure}[htb]
  \begin{center}
    \leavevmode
\epsfxsize=8.5cm
\epsffile{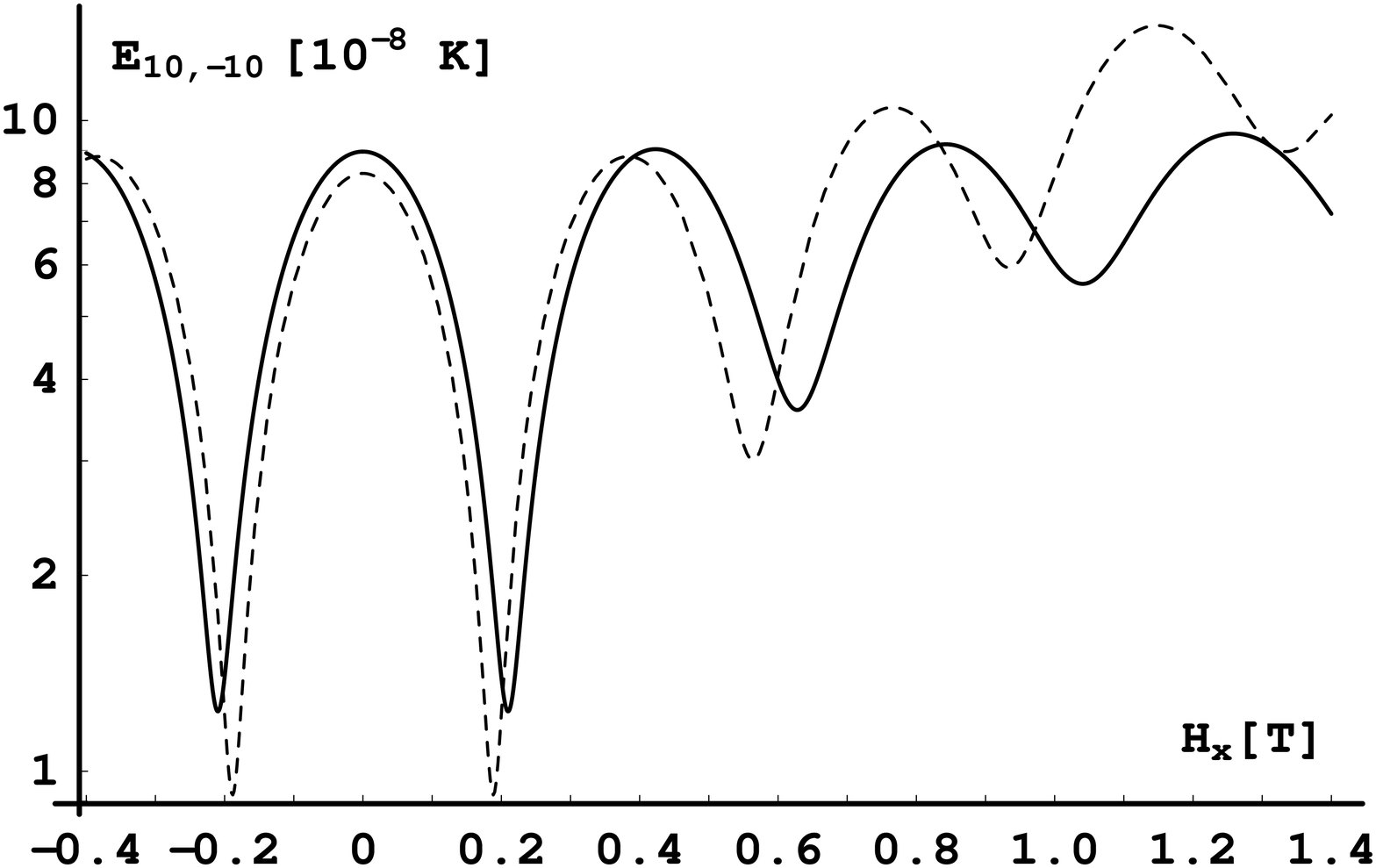}
  \end{center}
\caption{The tunnel splitting energy $E_{10,-10}(H_x)$ in Fe$_8$, exhibiting Berry phase oscillations, was calculated by exact diagonalization (solid line) and by using the approximate analytic formula (\protect\ref{gsplitting}) (dashed line). The angles are $\vartheta=90^\circ$, $\varphi=4^\circ$. The period is in excellent agreement with data (Refs.~\protect\onlinecite{Wernsdorfer} and \protect\onlinecite{Wernsdorfer2}) if we set $g=1.9$ (Ref.~\protect\onlinecite{g-factor}). The tunnel splittings for $H_x=0$ read: $E_{10,-10}=9.0\times 10^{-8}$ K, $E_{9,-9}=6.5\times 10^{-6}$ K, and $E_{8,-8}=2.1\times 10^{-4}$ K.}
\label{berry}
\end{figure}

In order to account for thermal transitions between the $\left|m\right>$ states, we include the most general spin-phonon coupling\cite{LeuenbergerLoss} which is allowed in leading order by the $D_2$ symmetry of the Fe$_8$ crystal,\cite{Wieghardt} i.e.,
\bea
\cH_{\rm sp} & = & g_1\epsilon_{xx}  S_x^2+g_2\epsilon_{yy}S_y^2
+\frac{1}{2}\left(g_3\epsilon_{xy}\left\{S_x,S_y\right\} \right.\nn\\
& & +\left.g_4\epsilon_{xz}\left\{S_x,S_z\right\}
+g_5\epsilon_{yz}\left\{S_y,S_z\right\} \right. \nn\\
& & +\left.g_6\omega_{xy}\left\{S_x,S_y\right\}
+g_7\omega_{xz}\left\{S_x,S_z\right\} \right. \nn\\
& & +\left.g_8\omega_{yz}\left\{S_y,S_z\right\}\right) ,
\label{H_sp}
\eea
where $g_i$, $i=1,\ldots,8$, are the spin-phonon coupling constants, which we assume to be approximately equal, $g_i\approx g_0$.\cite{LeuenbergerLoss} We know\cite{LeuenbergerLoss} that within the spin system the first- and second-order thermal transition rates $W_{m\pm 1,m}, W_{m\pm 2,m}$ are the strongest ones, which are evaluated by Fermi's golden rule to be ($r=1,2$)
\be
W_{m\pm r,m}=\frac{g_0^2s_{\pm r}}{q_r\pi\rho c^5\hbar^4}
\frac{(\varepsilon_{m\pm r}-\varepsilon_m)^3}
{e^{\beta(\varepsilon_{m\pm r}-\varepsilon_m)}-1}\, ,
\label{sp_rates}
\ee
where $s_{\pm 1}=(s\mp m)(s\pm m+1)(2m\pm 1)^2$, $s_{\pm 2}=(s\mp m)(s\pm m+1)(s\mp m-1)(s\pm m+2)$, and $q_1=48,q_2=32$. The mass density $\rho$ for Fe$_8$ is given by $1.92\times 10^3$ kg/m$^3$,\cite{Wieghardt} the sound velocity $c$ by 1400 m/s (yielding a Debye temperature of $\Theta_{\rm D}=33$ K).\cite{Gomes} The incoherent Zener probability $P_{\rm inc}$ in Fig.~\ref{zener_Fe8} fits the data\cite{Wernsdorfer2} well if one adjusts the coupling constant to $g_0=2.3$ K.

\begin{figure}[htb]
  \begin{center}
    \leavevmode
\epsfxsize=5cm
\epsffile{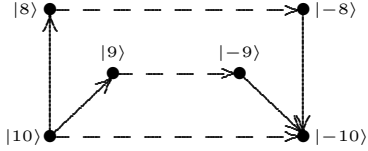}
  \end{center}
\caption{Unbiased ($n=0$) relaxation diagram for Fe$_8$. Solid (dashed) lines: thermal (tunneling) transitions.}
\label{diag_Fe8}
\end{figure} 

For the temperature range $0.05$ K$\le T\le 0.7$ K we achieve good agreement between our theory and the data\cite{Wernsdorfer2} if we take the states $\left|\pm 10\right>$, $\left|\pm 9\right>$, and $\left|\pm 8\right>$ into account. In particular, the path leading through $\left|\pm 8\right>$ gives a non-negligible contribution for $T\gtrsim 0.6$ K. Solving the relaxation diagram shown in Fig.~\ref{diag_Fe8} we obtain from Eq.~(\ref{rate_linear}) the following results for Fe$_8$ in the case $n=0$:
\bea
\Gamma_{\rm tot} & = & 2\left(\Gamma_{10}^{-10}+\sum\limits_{n=9}^8
\frac{b_n}{\frac{2}{W_{10,n}}+\frac{1}{\Gamma_n^{-n}}}\right), \\ 
\Delta\rho & = & \exp\left\{-\frac{\pi E_{10,-10}^2}{\hbar\alpha_{10}^{-10}}
-\sum\limits_{n=9}^8\frac{\pi E_{n,-n}^2W_{10,n}b_n}{\alpha_n^{-n}
\sqrt{E_{n,-n}^2+\hbar^2W_{10,n}^2}}\right\}, \nn
\eea
where we have used the approximation $\gamma_{n,-n}\approx W_{10,n}$ and $\left|\varepsilon_{mm'}^{<,>}\right|\gg E_{n,-n},\gamma_{n,-n}$.
$P_{\rm inc}=1-\Delta\rho$, which is plotted in Fig.~\ref{zener_Fe8}, is in good agreement with the measurements,\cite{Wernsdorfer2} except for the most narrow minima at $H_x\approx \pm 0.2$ T. However, the experimental uncertainty of these minima is very large since they depend strongly on the initial magnetization.\cite{Wernsdorfer2,Wernsdorfer_pc} In order to account for the increase of $P_{\rm inc}$ at $T=0.6,0.65,0.7$ K for higher fields $H_x$, we had to correct the energy levels $\varepsilon_m$ occurring in Eq.~(\ref{sp_rates}) by first-order perturbations in $H_x$. Below $0.4$ K, $P_{\rm inc}$ is $T$ independent, which agrees well with Ref.~\onlinecite{Wernsdorfer2}.

In conclusion, our theory, which is based only on thermal-assisted tunneling and neglects dipolar and hyperfine\cite{hyperfine} fields, agrees well with the recent measurements in Refs.~\onlinecite{Wernsdorfer} and \onlinecite{Wernsdorfer2} and also leads to the prediction of the anisotropies $B_4$, $B_6$, $B_8$, and the spin-phonon coupling constant $g_0$ in Fe$_8$.

Detailed calculations of the biased cases ($n=1,2,\ldots$) will be published elsewhere.

We are grateful to W. Wernsdorfer for providing us with unpublished data and for useful discussions. This work has been supported by the Swiss NSF.

\begin{figure}[htb]
  \begin{center}
    \leavevmode
\epsfxsize=8.5cm
\epsffile{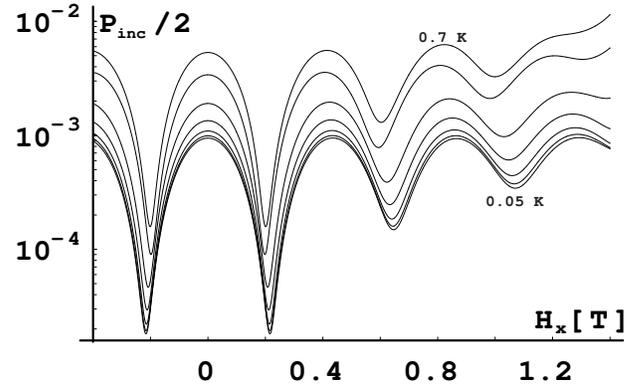}
  \end{center}
\caption{Zener transition probability $P_{\rm inc}(H_x)$ for temperatures $T=$0.7 K, 0.65 K, 0.6 K, 0.55 K, 0.5 K, 0.45 K, and 0.05 K. By choosing $B_4=-6.9\times 10^{-5}$ K for this plot our fit agrees well with data (Ref.~\protect\onlinecite{Wernsdorfer2}), except for the minima at $H_x\approx \pm 0.2$ T, which are too narrow (see text). Note that $P_{\rm inc}$ is equal to $2P$ in Ref.~\protect\onlinecite{Wernsdorfer2}. The tunnel splittings for this figure read: $E_{10,-10}=1.3\times 10^{-7}$ K, $E_{9,-9}=8.8\times 10^{-6}$ K, and $E_{8,-8}=2.7\times 10^{-4}$ K for $H_x=0$.}
\label{zener_Fe8}
\end{figure}



\begin{references}
\bibitem{g-factor} The value $g=1.9$ lies within experimental uncertainty of Ref.~\onlinecite{Barra}, since there only the anisotropies $A$ and $B_2$ are taken into account to obtain the fit value $g=2$.
\bibitem{hyperfine} Hyperfine effects in Fe$_8$ become noticeable for $T\lesssim 1.5$ K, and they lead to measurable corrections to $P_{\rm inc}$, which are small in the sense that the overall shape of the curve $P_{\rm inc}(T)$ is well described by spin-phonon effects (Ref.~\onlinecite{Wernsdorfer2}). Also we note that the value for $E_{10,-10}$ reported for
$^{\rm st}$Fe$_8$ and for $^{57}$Fe$_8$ differ by less than 20\% (Ref.~\onlinecite{Wernsdorfer2}), whereas the value reported initially for $^{\rm st}$Fe$_8$ (Ref.~\onlinecite{Wernsdorfer}) and the one reported subsequently for $^{\rm st}$Fe$_8$ (Ref.~\onlinecite{Wernsdorfer2}) differ by more than 25\%. Given this experimental uncertainty further measurements would be desirable.
\end{references}
\end{document}